\documentclass[aps,prd,10pt,twocolumn,preprintnumbers,superscriptaddress,showpacs,floatfix,nofootinbib]{revtex4-1}
\usepackage{textcomp,gensymb}
\usepackage{hyperref}
\usepackage{graphicx}
\usepackage{amsfonts,amsmath,amssymb,bm,bbm}
\usepackage{color,soul}
\usepackage{xcolor}
\usepackage{slashed}
\usepackage[toc,page]{appendix}
\usepackage{flushend}
\usepackage{physics}
\usepackage{graphicx}
\usepackage{dcolumn}
\usepackage{float}
\usepackage[toc,page]{appendix}

\newcommand{\mnras}{MNRAS}

\hypersetup{
    pdfnewwindow=true,      
    colorlinks=true,       
    linkcolor=blue,          
    citecolor=blue,        
    filecolor=blue,      
    urlcolor=blue        
}

\begin{document}

\title{Can a Dark Inferno Melt Earth's Core?}

\author{Christopher Cappiello}
\affiliation{Department of Physics and McDonnell Center for the Space Sciences, Washington University, St. Louis, MO 63130, USA}

\author{Tansu Daylan}
\affiliation{Department of Physics and McDonnell Center for the Space Sciences, Washington University, St. Louis, MO 63130, USA}

\begin{abstract}
The search for dark matter is one of the crucial open problems in both particle physics and cosmology. If dark matter scatters with Standard Model particles, it could accumulate inside the Earth and begin to annihilate, producing heat within the Earth's core. While past work has been done on the effect that this heat would have once it reached the surface, we model the flow of heat through the Earth's core by numerically solving the heat equation to model dark matter's effect on the interior of the planet. We compute how long it takes for the core to come into thermal equilibrium and show that for a wide range of dark matter parameters, a substantial fraction of the inner core would be melted by dark matter annihilation. Our analysis produces new limits on dark matter annihilating in the Earth, points out important new effects that must be considered when studying planetary heating by dark matter, and suggests new dark matter observables that could be searched for in exoplanet populations.
\end{abstract}

\maketitle

\section{Introduction}
Indirect detection, one of the central ways of searching for dark matter (DM), is largely based on the principle that dark matter may decay or annihilate into Standard Model particles (see e.g.~\cite{Buckley:2013bha}). Because the annihilation rate scales with the dark matter density squared, any process that concentrates dark matter to a high density can increase the overall annihilation rate, as well as provide clear locations to search for annihilation products. For this reason, searches for dark matter annihilation often focus on locations with high expected dark matter equilibrium density, such as the Galactic Center~\cite{Hooper:2010mq,Daylan2016,ANTARES:2019svn,MAGIC:2022acl,HESS:2022ygk,veritas,Foster:2022nva,IceCube:2023ies} and dwarf galaxies~\cite{Kerszberg:2023cup,McGrath:2023oto,HAWC:2017mfa,Geringer-Sameth:2014qqa,Fermi-LAT:2011vow,Geringer-Sameth:2011wse,Fermi-LAT:2010cni,Strigari:2018utn}. However, it is also possible for dark matter to be captured in celestial objects with high baryonic density (e.g. stars and planets), build up inside them, and eventually begin to annihilate.

If dark matter has a nonzero scattering cross section with nucleons or electrons, then it should occasionally collide with stars, planets, and other celestial bodies. If a dark matter particle loses enough energy in such a collision, it can become gravitationally captured, while crucially remaining on an orbit that continuously passes through the object. Over millions or billions of years, it can continue scattering, losing more energy and gradually sinking toward the object's center. A sizable density is thus built up, which can begin to annihilate. If the dark matter annihilates to neutrinos~\cite{Freese:1985qw,Silk:1985ax,Gaisser:1986ha,Griest:1986yu,Srednicki:1986vj,Super-Kamiokande:2004pou,Super-Kamiokande:2011wjy,ANTARES:2016xuh,Super-Kamiokande:2015xms,IceCube:2016dgk,Liu:2020ckq,Maity:2023rez,IceCube:2021xzo}, or to long-lived mediators that can escape the celestial body~\cite{Batell:2009zp,Schuster:2009fc,Bell:2011sn,Feng:2016ijc,ANTARES:2016obx,Leane:2017vag,HAWC:2018szf,Bell:2021pyy,Nguyen:2022zwb,Linden:2024uph}, then the products of this annihilation or of the subsequent decay chains can be detected on Earth.

\begin{figure}
    \centering
    \includegraphics[width=\linewidth]{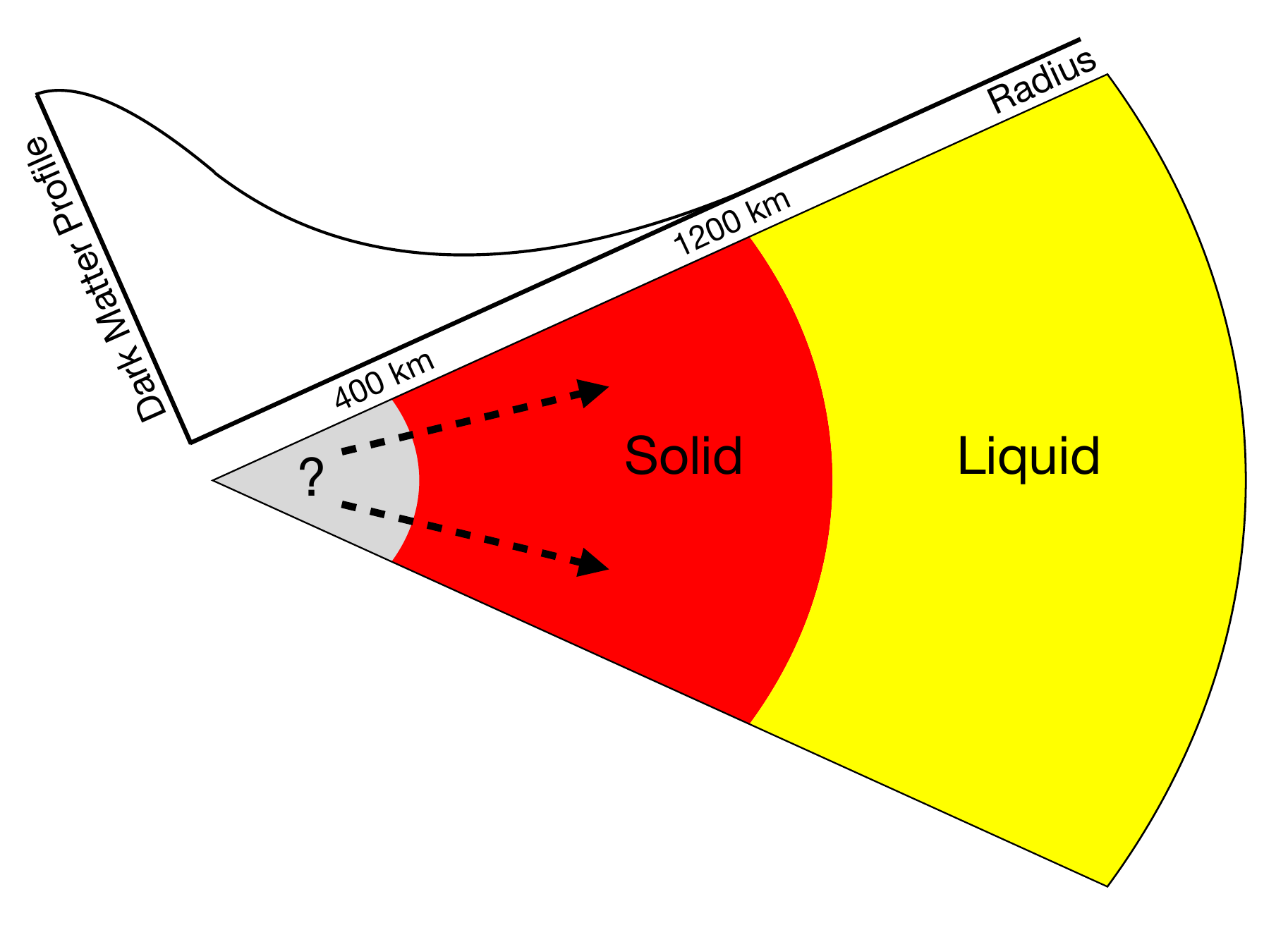}
    \caption{Diagram of the Earth's core, with an example dark matter density profile superimposed. The solid-liquid boundary at 1200 km is the boundary between the inner core and outer core, while a radius of 400 km is where the seismic data we use loses sensitivity to the core's structure (see text for details). Dashed arrows denote heat flow away from the center of the Earth.}
    \label{fig:earthdiagram}
\end{figure}

If, on the other hand, dark matter annihilates to photons or charged particles, those particles will not escape, and will instead contribute to heating the star or planet in question (see Fig.~\ref{fig:earthdiagram}). This can manifest in several different observables, including affecting the measured heat flow of the Earth~\cite{Mack:2007xj,2013JPhG...40k5202M,Bramante:2019fhi}, the surface temperature of exoplanets~\cite{Leane:2020wob,Acevedo:2024zkg,Benito:2024yki}, emission from planetary atmospheres~\cite{Blanco:2023qgi,Blanco:2024lqw}, and planet or star formation itself~\cite{Croon:2023bmu,John:2023knt}. These ideas have been used to set limits on dark matter ranging from the MeV or GeV scale (depending on the treatment of dark matter evaporation) up to roughly $10^{10}$ GeV.

These analyses typically begin by assuming that the system is in equilibrium, i.e., that the dark matter capture rate equals the annihilation rate, and that the energy injection rate due to annihilation equals the heat flow near the surface. However, previous works have largely neglected the effect that this heating would have on the interior of a planet. Nonzero heat flow requires a nonzero temperature gradient, and given that the annihilation rate quite generically peaks in the center of a planet, this annihilation should heat the core of a planet more than the surface. In this work, we show that, in the parameter space previously constrained by Earth's heat flow, dark matter annihilation would have melted a substantial portion of Earth's inner core, which conflicts with seismic analyses of the inner core's structure~\cite{Pang23}. We thus set limits that are about an order of magnitude stronger than previous heat flow limits over a broad mass range. We model the flow of heat within the Earth over time, and show that the time for Earth to come into thermal equilibrium can be significant on geological scales. We also emphasize more generally the impact that dark matter can have on planetary structure, which could affect exoplanet magnetic fields and alter how we understand dark matter's effects on planet formation. 

\section{Dark Matter Heating of Earth}

\subsection{Dark Matter Capture in the Earth}

The study of DM capture in the Earth and/or Sun goes back to the 1980's~\cite{1985ApJ...296..679P,1985ApJ...299..994F, 1987ApJ...321..571G}. When DM particles scatter with nuclei or electrons in a planet or star, they may lose enough energy to be gravitationally captured. Once captured, they will continue to scatter and, given enough time, may thermalize within the planet/star. For DM with a mass comparable to the masses of the target nuclei, capture may be possible with a single collision (see e.g.~\cite{1987ApJ...321..571G}). On the other hand, DM much heavier than nuclei typically requires many collisions in order to be sufficiently slowed~\cite{Bramante:2017xlb}. DM lighter than about the GeV scale is expected to evaporate from many celestial objects, i.e., its thermal velocity would be above escape velocity, suppressing the capture rate~\cite{Garani:2021feo}; this effect can be evaded in some models, see e.g. Ref.~\cite{Acevedo:2023owd}.

To compute the capture rate within the Earth, we use the Python package \textsc{Asteria}~\cite{rebecca_k_leane_2023_8150110}, developed by the authors of Ref.~\cite{Leane:2023woh}. We use the same parameters for the Earth provided in the public distribution of \textsc{Asteria}, including a local DM density of 0.4 GeV/cm$^3$ and a mean incoming velocity of 270 km/s, which reproduce the capture rates shown in Ref.~\cite{Leane:2023woh}.  While we do not derive the capture rate within the Earth here, we direct the reader to a series of recent papers presenting both analytic and simulation-based treatments of DM capture~\cite{Bramante:2017xlb,Ilie:2021umw,Bramante:2022pmn,Leane:2023woh,Ilie:2024sos}.

\subsection{Earth's Core: Physical Parameters}

The Earth's core is, of course, difficult to study directly. As a result, many of its properties must be inferred from laboratory analogs and/or theoretical calculations. For example, efforts have been made to measure or predict the thermal conductivity of various iron alloys at extremely high temperatures and pressures, but these have resulted in values ranging from $\sim$30 W/m/K~\citep{Konopkova16} to over 200 W/m/K~\citep{Pozzo14}. We adopt an intermediate value of 100 W/m/K, consistent with the measurement reported in Ref.~\cite{OhtaConductivity} (see also Refs.~\cite{DeKokerConductivity,2013PEPI..224...88G}, which give values around 150 W/m/K).

Similarly, the temperature of Earth's inner core is subject to uncertainties at the $\mathcal{O}$(10$\%$) level: reported temperatures of the inner core boundary (ICB) range from 4500$\sim$6500 K, with the very center of Earth being a couple hundred Kelvin higher (see Refs.~\cite{1987Sci...236..181W,2002physics...4055S,doi:10.1126/science.1233514} and references therein). We adopt an intermediate value of 5500 K.

\subsection{Dark Matter Heating}

The form of the DM density profile within the Earth depends on the strength of the DM's interactions with Standard Model particles. When the interaction is strong and the mean free path short, DM can be treated as being in local thermal equilibrium, as was done in Ref.~\cite{Gould:1989hm}. In this work, we consider relatively weak interactions, such that the mean free path is typically at least comparable to the scale radius of the DM distribution. We therefore work in the isothermal regime, which was treated in Ref.~\cite{1985ApJ...294..663S}. In this limit, the DM density profile is given by~\cite{Vincent:2015gqa}

\begin{equation}
    n_{\chi}(r) = n_0\frac{e^{-\frac{r^2}{r_{\chi}^2}}}{\pi^{3/2}r_{\chi}^3}\,,
\end{equation}
where $n_0$ is the total number of DM particles and $r_{\chi}$ is a scale radius given by

\begin{equation}
    r_{\chi} = \sqrt{\frac{3 k_B T_c}{2 \pi G \rho_c m_{\chi}}}\,.
\end{equation}
Here $T_c$ is the core temperature, $\rho_c$ is the core density, $k_B$ is Boltzmann's constant, and $G$ is the gravitational constant.

The annihilation rate is simply proportional to the DM density squared. We assume that capture and annihilation of DM are in equilibrium, i.e. $\Gamma_{cap} = \dot{Q}_{\chi}$, where $\Gamma_{cap}$ is the mass capture rate and $\dot{Q}_{\chi}$ is the total dark matter heat injection. This assumption may not be valid if the annihilation cross section is too small. However, this assumption is satisfied for weak-scale cross sections, assuming s-wave annihilation~\cite{Mack:2007xj,Bramante:2019fhi}. We furthermore assume that all of the DM mass energy is converted into heat, which is a reasonable assumption as long as the DM is much heavier than its annihilation products (or their eventual decay products); as shown below, the DM masses we consider are at least 10 GeV, making this a well justified assumption.

We model heat flow due to DM using the heat equation in spherical coordinates. Assuming spherical symmetry (i.e., that the temperature profile has no dependence on $\theta$ or $\phi$), this equation is
\begin{equation}
    \frac{1}{\alpha}\frac{\textrm{d}u}{\textrm{d}t} = \frac{\textrm{d}^2u}{\textrm{d}r^2} + \frac{2}{r}\frac{\textrm{d}u}{\textrm{d}r} + \frac{1}{k}\dot{q}\,.
\end{equation}
Here $k$ is the thermal conductivity, and $\alpha$ is the thermal diffusivity, given by $\alpha = \frac{k}{c\rho}$, where $c$ is the specific heat capacity and $\rho$ is the mass density of the material. As mentioned above, we use a fiducial value of $k = 100$ W/m/K, though we will vary this value to study its impact on our conclusions, and a value of $\alpha = 24 \times 10^{-6}$ m$^2$/s~\cite{thermaldiffusivity}. $\dot{q} \propto n_{\chi}(r)^2$ is the volumetric DM heat injection rate, with a normalization set by the capture rate. 

To solve this equation, we assume that the inner core initially has a constant temperature $T_0$, that the temperature at the inner core boundary is $T_0$ at all times, and that the radial derivative of the temperature profile at $r=0$ is zero. For most of our results, we set $T_0 = 5500$ K, but we will vary this value to study how a different value would affect our conclusions.

\begin{figure}[t]
    \centering
    \includegraphics[width=\linewidth]{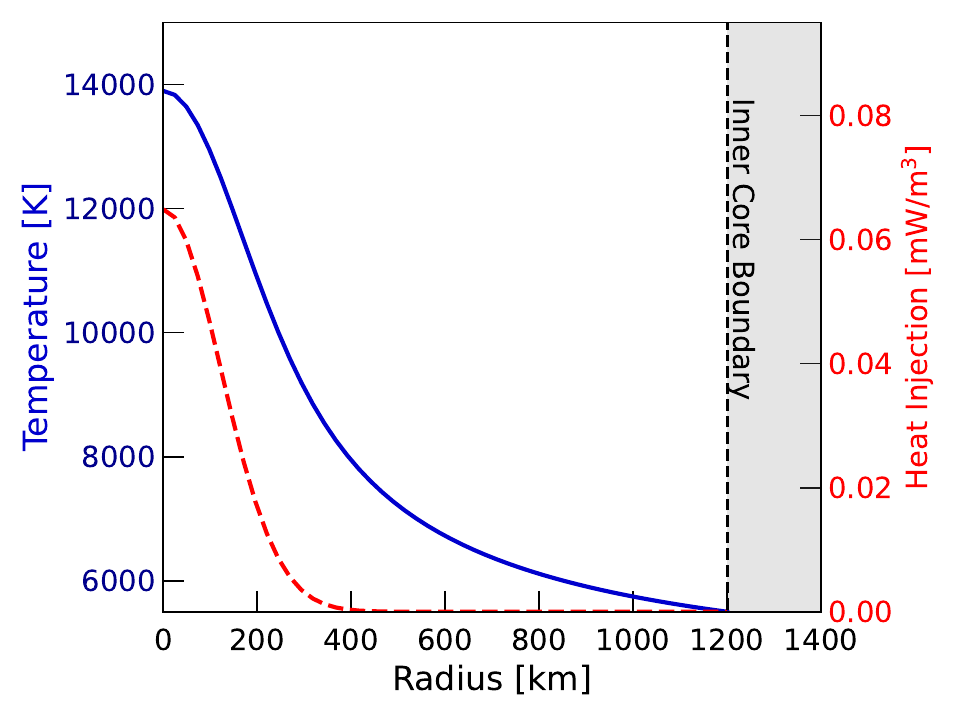}
    \caption{Steady-state temperature (solid blue) and heat injection (dashed red) profiles for $m_{\chi} = 1$ TeV and $\sigma_{\chi N} = 10^{-36.9}$ cm$^2$, assuming $k = 100$ W/m/K and $T_0 = 5500$ K. Note that the temperature axis begins at 5500 K.}
    \label{fig:tempandheatinjection}
\end{figure}

Figure~\ref{fig:tempandheatinjection} shows the steady-state temperature profile and heat injection rate, as a function of radius, for $m_{\chi} = 1$ TeV and $\sigma_{\chi N} = 10^{-36.9}$ cm$^2$. We only solve for these quantities out to the inner core boundary at 1200 km; the extension of the x-axis is purely to make the position of this boundary apparent. Note that for this value of $m_{\chi}$, virtually all of the heat is deposited deep within the inner core.

\begin{figure}[t]
    \centering
    \includegraphics[width=\linewidth]{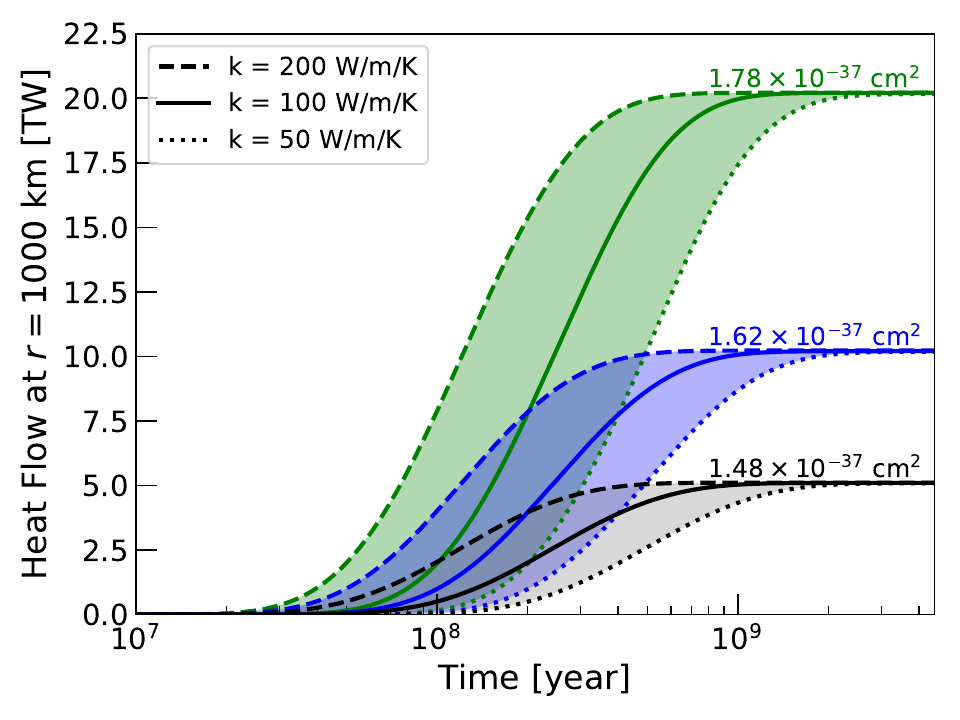}
    \caption{Dark matter-sourced heat flow through a spherical shell of radius 1000 km as a function of time. All curves assume $m_{\chi}$ = 1 TeV and a uniform starting temperature of 5500 K for the inner core. We show results for 3 values of the inner core's thermal conductivity $k$ (shown in different line styles), and 3 values of the cross section $\sigma_{\chi N}$ (in different colors).}
    \label{fig:heatflowvstime}
\end{figure}

If heat is deposited deep within the Earth's core, which is the case for DM heavier than about 10 GeV and a temperature of 5500 K, it will take a finite time for that heat to be conducted toward the surface of the planet. Figure~\ref{fig:heatflowvstime} shows the amount of DM-produced heat flowing out through a shell of radius 1000 km, assuming a DM mass of 1 TeV and various choices for the cross section $\sigma_{\chi N}$. The values of $\sigma_{\chi N}$ used are chosen to correspond to $\dot{Q}_{\chi} = $\{5 TW, 10 TW, 20 TW\}. We see that, depending on the value of $k$, but roughly independently of the power injected, it can take of order $10^9$ years for the full power injected to begin flowing through a radius of 1000 km.

Although this timescale of $10^9$ years is valid for setting limits based on the properties of the Earth, this may not be true for other (exo)planetary observables. Convection in the liquid cores of planets, or in the atmospheres of gas giants, would transfer heat much more efficiently than the conduction we consider. Several references have proposed using exoplanets as DM detectors~\cite{Leane:2020wob,Croon:2023bmu,Acevedo:2024zkg,Benito:2024yki}. Planets near the Galactic Center (where the DM density is high) would capture much more DM than similar planets in the Solar System. In contrast, planets with particularly low temperatures would be sensitive to quite small DM heating rates. The very formation of gas giants has even been used to set constraints on DM properties. In setting constraints based on young planets, it is important to be sure not only that the capture and annihilation rates have time to equilibrate, but also that the planet's temperature profile has come into equilibrium as well.

\subsection{Melting the Earth's Core}

Here, we derive a limit on the DM-proton cross section based on the requirement that DM annihilation not melt Earth's inner core. We begin this Section by noting that the Earth's core is not as old as the planet itself, though estimates of its age differ: Ref.~\cite{OhtaConductivity} reports an age of less than 700 Myr~\cite{OhtaConductivity}, and Ref.~\cite{PhysRevLett.125.078501} finds a value of 1--1.3 Gyr, while Ref.~\cite{Konopkova16} allows for a solid inner core as old as 4.2 Gyr. However, this does not affect our conclusions, for the following reason: if DM produces enough heat to melt the core from solid and subsequently keep it liquid, then it certainly produces enough to keep it in liquid form if it was already molten.

\begin{figure}[t]
    \centering
    \includegraphics[width=\linewidth]{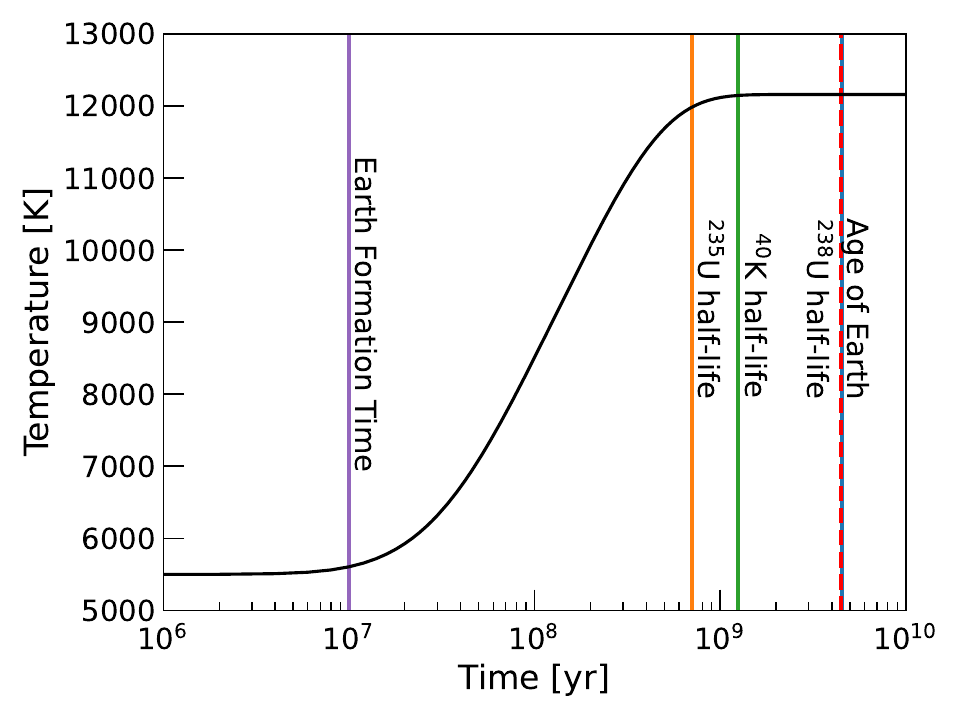}
    \caption{Temperature at $r$ = 400 km as a function of time, assuming 5 TW power injection and $m_{\chi}$ = 1 TeV. For comparison, we show the half-lives of several nuclei that contribute to radiogenic heating of the Earth, as well as the age of the Earth, and an approximate estimate of the time it took for Earth to form.}
    \label{fig:tempvstime}
\end{figure}

As mentioned above, seismic data provide measurements of the porosity of the inner core down to quite large depths. Kilometer-scale inhomogeneitites in the inner core scatter high frequency seismic waves, producing a time-delayed signal that can be used to study the structure of these inhomogeneities. We refer to Ref.~\cite{Pang23}, which provides measurements of the heterogeneity of the inner core down to depths of 800 km below the inner core boundary (about 400 km from the center of the Earth), losing resolution at lower depths. In setting a limit, we very conservatively allow that DM could have melted the inner 400 km of the inner core. However, note that extending this melted region any further would conflict with seismic data. We also noted above that there is a disagreement of thousands of K on the melting point of iron (or various iron alloys) at core conditions. To be extremely conservative, we assume that any inner core material that reaches $10^4$ K would be fully melted (compare to the melting point of 7600 $\pm$ 500 K for iron at inner core boundary conditions reported by Ref.~\cite{1987Sci...236..181W}). Following both of these assumptions, we set a simple and conservative requirement: DM is ruled out if it would cause the temperature to rise to at least $10^4$ K throughout a sphere of radius $r_{melt} > 400$ km centered on the center of the Earth.

In Fig.~\ref{fig:tempvstime}, we show the temperature at a radius of 400 km as a function of time, assuming $m_{\chi}$ = 1 TeV and $\dot{Q}_{\chi}$ = 5 TW. For reference, we also mark the age of the Earth, the half-lives of several isotopes that contribute to radiogenic heating, and a rough estimate of the time it took Earth to form. This plot is similar in shape to the curves shown in Fig.~\ref{fig:heatflowvstime}, with the temperature initially constant, then rising to an equilibrium value (in this case, around 12,000 K). We see that the time for this equilibrium temperature to be reached is again around 1 Gyr, well shorter than the age of the Earth. Given this plot, we can conclude that any value of the DM-nucleon cross section $\sigma_{\chi N}$ that produces 5 TW of power for $m_{\chi} = 1$ TeV is ruled out.

\subsection{\textsc{DarkInferno}}

We developed our computational framework into a Python package, \textsc{DarkInferno}\footnote{https://github.com/AstroMusers/darkinferno}\cite{chris_cappiello_2025_15530200}, which is designed to make our results reproducible, optimize the computational efficiency of our model evaluations, provide flexible functionality to generalize our differential equation solutions to other rocky planets in the Solar system and exoplanets. \textsc{DarkInferno} provides solutions both for the time-independent heat equation, assuming that the core has reached thermal equilibrium, and the full time-dependent heat equation using a finite difference method. In what follows, we use the time-dependent method to determine the time it takes for the Earth to come to thermal equilibrium, and once we have established that this is short compared to the age of the planet, we use the much more efficient time-independent approach to determine the steady-state temperature profile of the inner core. 
To compute the capture rate in Earth, \textsc{DarkInferno} uses the \textsc{Asteria} Python package, with the default inputs for Earth capture as discussed above.

\section{Results}

\subsection{Limits}

\begin{figure}[t]
    \centering
    \includegraphics[width=\linewidth]{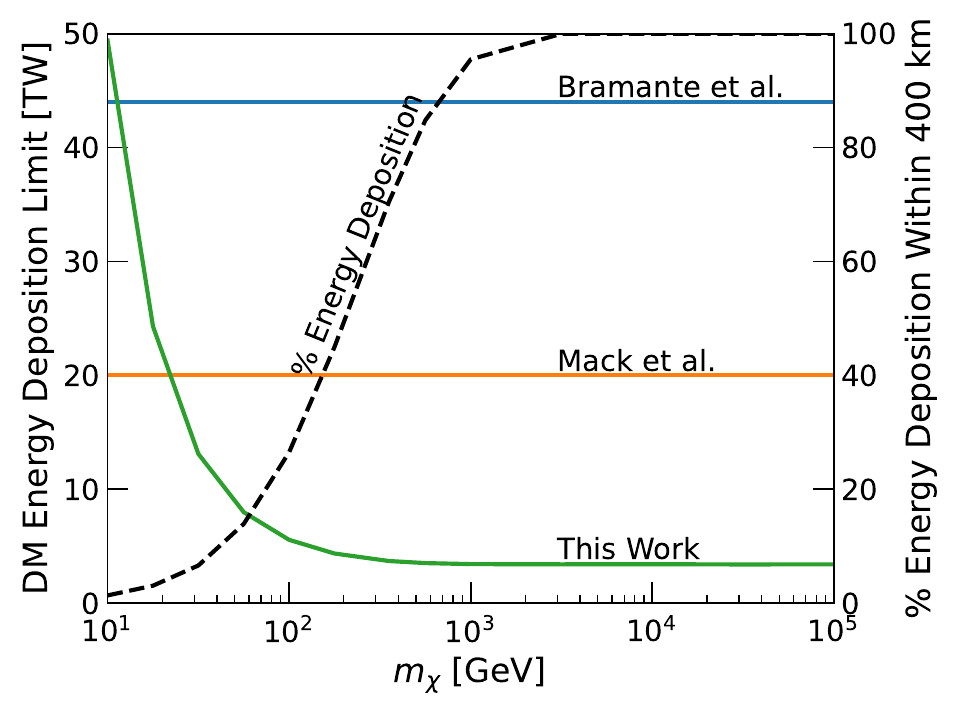}
    \caption{DM power injection required to set a limit, as a function of the DM mass $m_{\chi}$. We show the constant values of 44 TW (blue, from Ref.~\cite{Bramante:2019fhi,Bramante:2022pmn}) and 20 TW (orange, from Ref.~\cite{Mack:2007xj}) used in previous literature, along with our limit (green). Shown in dashed black is the percentage of energy that is injected within the innermost 400 km (labeled on the right-hand vertical axis).}
    \label{fig:energydepositionlimit}
\end{figure}

In Fig.~\ref{fig:energydepositionlimit}, we show the amount of DM energy injection required to melt the core out to 400 km, assuming $k$ = 100 W/m/K and $T_0$ = 5500 K. Past work has set limits on DM annihilating within the Earth by requiring that it not account for more than 20 TW~\cite{Mack:2007xj}, or the full observed 44 TW~\cite{Bramante:2019fhi,Bramante:2022pmn}, of heat flow from within the Earth; we show both these values for comparison. When DM is lighter than about 100 GeV, the scale radius of the DM distribution $r_{\chi}$ is larger than 400 km, meaning that some of the DM annihilation occurs at larger radii and has little effect on the temperature in the innermost radii. This is especially true at 10 GeV, for which $r_{\chi}$ is larger than the radius of the inner core. For larger $m_{\chi}$, effectively all the annihilation occurs within the innermost 400 km. Once this is the case, increasing $m_{\chi}$ further has essentially no impact on the temperature at 400 km, resulting in the curve flattening off. At large masses, our limiting heat injection reaches 3.4 TW, a factor of 6--13 less than was required in previous analyses. In this sense, considering the structure of the inner core is around an order of magnitude more sensitive than simply comparing the DM signal with the Earth's internal heat flow.

\begin{figure}[t]
    \centering
    \includegraphics[width=\linewidth]{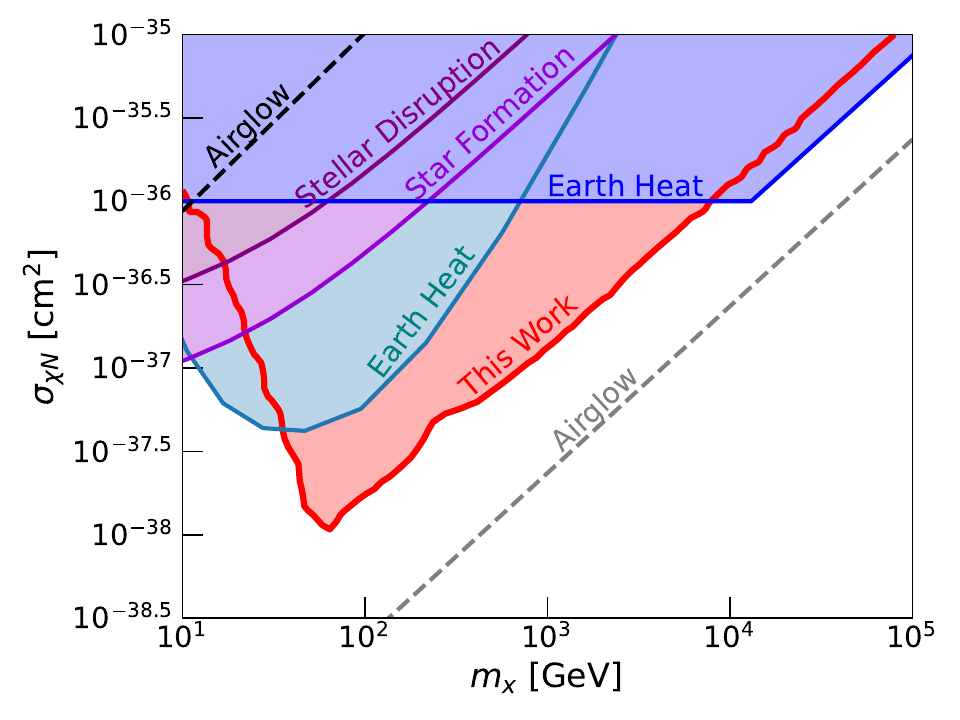}
    \caption{Our limit from the existence of a solid inner core is shown in red. We also show other limits from Earth heating (teal and blue)~\cite{Mack:2007xj,Bramante:2019fhi,Bramante:2022pmn}, star formation and stellar disruption (purple and burgundy, respectively)~\cite{John:2023knt}, and airglow in gas giant atmospheres (dashed black and dashed gray)~\cite{Blanco:2023qgi,Blanco:2024lqw}.}
    \label{fig:limit}
\end{figure}

As long as capture and annihilation are in equilibrium, the heat deposition rate equals the mass capture rate, and thus the limit on energy deposition can be converted into a limit on $\sigma_{\chi N}$. Figure~\ref{fig:limit} shows our limit on $\sigma_{\chi N}$ as a function of $m_{\chi}$. At low mass, the limit becomes weaker because, as discussed above, most of the energy injection does not occur in the innermost 400 km of the core (see Fig.~\ref{fig:energydepositionlimit}). At large mass, the DM is much heavier than any nucleus in the Earth, and must scatter many times in order to be slowed appreciably, thus requiring a larger cross section to be captured efficiently.

For comparison, we show several other limits that arise from considering DM capture in planets and stars. The teal line is the Earth heating limit from Ref.~\cite{Mack:2007xj}. The blue limit combines two limits from Earth heating, from Refs.~\cite{Bramante:2019fhi,Bramante:2022pmn}. The limit becomes flat at $\sigma_{\chi N} = 10^{-36}$ cm$^2$ because, at lower cross sections, the assumption of local thermal equilibrium is no longer valid, and the authors of Ref.~\cite{Bramante:2022pmn} cut off their limit at that point. Their limit is stronger than ours at large cross sections, but this is due to them computing a larger capture rate than we obtain from \textsc{Asteria}.

The purple and burgundy curves are limits that arise from the effect of DM capture on star formation and stellar evolution, respectively~\cite{John:2023knt}. The dashed gray line is a limit set based on the fact that DM annihilation could cause molecular excitation and resulting ultraviolet emission in the atmospheres of Solar System gas giants~\cite{Blanco:2024lqw}; similarly, the dashed black curve is based on a search for the analogous production of ionized H$_3$ on Jupiter. These latter limits are shown as dashed lines because, for the annihilation products to reach the planetary atmosphere, the analyses require that DM annihilate via a light mediator that escapes the core of the planet and decays in the atmosphere. Thus, the model in question is different from the one constrained here, and bypasses the considerations of heat transfer and internal heating that are central to the present work.

\section{Conclusions}

In this work, we have shown that dark matter annihilating inside the Earth could heat the core enough to melt it. Despite the extreme nature of this effect, the amount of dark matter annihilation required to accomplish it is substantially less than can be ruled out from considerations of terrestrial heat flow, as has been done in previous work. We thus set bounds on dark matter's spin-independent scattering cross section, which are stronger than recent bounds based on dark matter capture in the Earth and stars.

We acknowledge that the parameter space we constrain is already ruled out by direct detection experiments. However, in this work, we have identified a novel signature that could be applied to a variety of planets, as well as a variety of dark matter models. Past work has proposed to look for signs of dark matter capture in particularly massive planets, which could efficiently capture dark matter that is light enough to be unconstrained by direct detection~\cite{Leane:2024bvh}. Recent work has also shown that additional long-range interactions could bind dark matter to a planet, similarly preventing evaporation and allowing sensitivity to quite low masses~\cite{Acevedo:2023owd}. On the other hand, by modeling the dark matter density in the local thermal equilibrium regime, our work could be extended to larger cross sections, where direct detection experiments lose sensitivity due to attenuation in the Earth~\cite{Kavanagh:2017cru,Emken:2018run,Cappiello:2023hza}.

At the same time, we have shown that the time it takes heat injected in the center of a planet to reach the inner core boundary---to say nothing of the planetary surface---can be significant on geological scales. This consideration is of crucial importance when setting limits on dark matter based on planet formation, or on observations of exoplanets that may be much younger than the Earth~\cite{2018haex.bookE.184C,2023MNRAS.520.5283G,2024ApJ...976..234B}.

\acknowledgments

T.D. acknowledges support from the McDonnell Center for the Space Sciences at Washington University in St. Louis. CVC was generously supported by Washington University in St. Louis through the Edwin Thompson Jaynes Postdoctoral Fellowship.

We are grateful to Rebecca Leane for updates to the \textsc{Asteria} dark matter capture code, and to both Rebecca Leane and Marianne Moore for help using this code. We also thank Isabelle John and Marianne Moore for providing extrapolations of their constraints to larger masses. Finally, we thank John Beacom, Joe Bramante, Djuna Croon, Juri Smirnov, and Aaron Vincent for helpful comments on the manuscript.

\clearpage

\appendix
\section{Uncertainties}

As noted above, there are significant uncertainties in both the initial temperature of the inner core $T_0$ and the thermal conductivity $k$ within the inner core. To quantify the impact of these uncertainties on our limit, we computed the value of the limit at $m_{\chi} = 1$ TeV for several hundred combinations of $k$ and $T_0$. We considered 31 values of $T_0$ spaced linearly from 4000 K to 7000 K, and 22 values of $k$, sampled in log space from 20 to 224 W/m/K, resulting in a total of 682 points.

\begin{figure}[H]
    \centering
    \includegraphics[width=\linewidth]{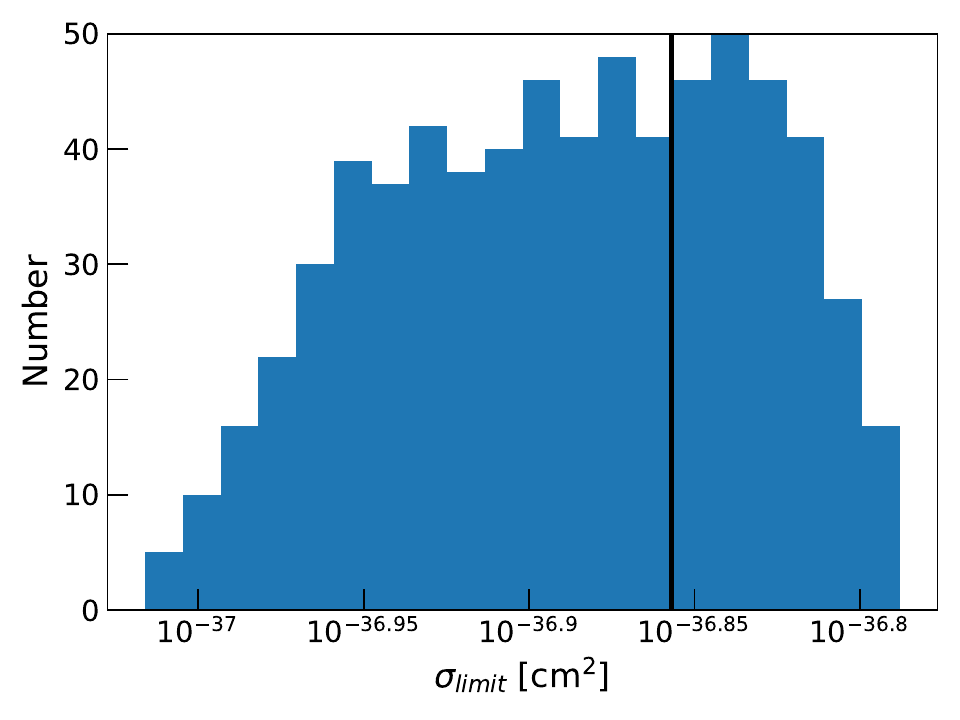}
    \caption{Histogram of values obtained for the limit at $m_{\chi} = 1$ TeV, for numerous combinations of the inner core's thermal conductivity and initial temperature. See text for details. The black line is the value of the limit for $k = 100$ W/m/K and $T_0$ = 5500 K.}
    \label{fig:histogram}
\end{figure}

\begin{figure}[H]
    \centering
    \includegraphics[width=\linewidth]{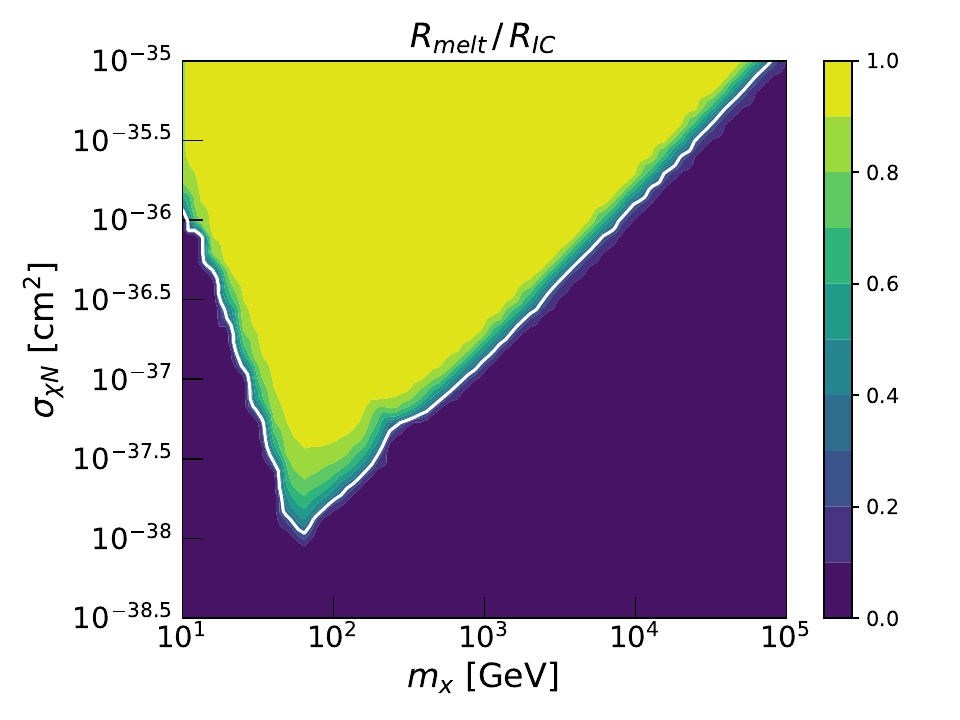}
    \caption{Radius out to which the inner core is melted, as a fraction of the total radius of the inner core, for different values of $m_{\chi}$ and $\sigma_{\chi N}$. Here we assume $k$ = 100 W/m/K and $T_0$ = 5500 K. Dark blue denotes little to no melting, while yellow denotes approximately the entire inner core being melted. Our limit, where $r_{melt}$ = 400 km, is shown as the white curve.}
    \label{fig:parameterspace}
\end{figure}

Figure~\ref{fig:histogram} shows a histogram of the various values of our limit at $m_{\chi} = 1$ TeV, which we obtain by scanning over $k$ and $T_0$ as described above. The black line is the limit we obtain for $k = 100$ W/m/K and $T_0$ = 5500 K, the values assumed in Fig.~\ref{fig:parameterspace}. We see that the entire range of variation in the limit is a factor of $\sim$1.6, suggesting that the overall range of values for our limit is less than $\pm$ 30\%.

Our limit could also shift if we required a larger or smaller fraction of the inner core to be melted in order to rule out DM. In Fig.~\ref{fig:parameterspace}, we show the radius out to which the inner core is melted, across the parameter space of $\sigma_{\chi N}$ vs. $m_{\chi}$. This figure assumes that $k$ = 100 W/m/K and $T_0$ = 5500 K, as in Fig.~\ref{fig:limit} Dark blue denotes little to no melting, while yellow shows where essentially the entire inner core would be melted. We see that the range of possible cross section limits set by requiring different fractions of the inner core to melt is only about a factor of 3.

\bibliography{main}

\end{document}